\documentclass{article}
\usepackage{amssymb, amsmath}
\numberwithin{equation}{section}
\newtheorem{theorem}{Theorem}[section]
\newtheorem{proposition}{Proposition}[section]

\newtheorem{remark}{Remark}[section]
\newtheorem{definition}{Definition}[section]
\begin{document}
\title{Global existence of solutions for the relativistic Boltzmann equation on
 the flat Robertson-Walker space-time for arbitrarily large initial data}
 \author{Norbert NOUTCHEGUEME$^{1}$; Etienne TAKOU$^{2}$ \\
$^{1}$Department of Mathematics, Faculty of Sciences, University
of Yaounde 1,\\ PO Box 812, Yaounde, Cameroon \\ {e-mail:
nnoutch@justice.com$^{1}$,nnoutch@uycdc.uninet.cm$^{1}$,
etakou@uycdc.uninet.cm$^{2}$}}
\date{}
\maketitle
\begin{abstract}
We prove a global in time existence and uniqueness theorem for the
initial value problem for the relativistic Boltzmann equation on
the flat Robertson- Walker space-time, in a functional framework
appropriate to the coupling with Einstein's equations. We had
nowhere to restrict the size of the initial data which can hence
be taken arbitrarily large.
\end{abstract}
\section{Introduction}
The Boltzmann equation is one of the basic equations of the
kinetic theory. This equation rules the dynamics of one kind of
particles subject to mutual collisions, by determining their
distribution function, which is a non-negative real-valued
function of both the position and the momentum of the particles.
The distribution function is physically interpreted as the
"probability of the presence density" of the particles in a giving
volume during their collisional evolution.

In the case of binary and elastic collisions, the distribution
function is determined by the Boltzmann equation through a
non-linear operator called the "Collisions operator" , that
describes, at each point where two particles collide with each
other, the effects of the behaviour imposed by the collision to
the distribution function, taking into account the fact that the
momentum of each particle is not the same, before and after the
collision, only the sum of their two momenta being preserved.

Due to its importance in the kinetic theory, several authors
studied and proved local and global existence theorems for the
Boltzmann equation, in both the \textbf{non-relativistic} case,
that considers particles with low velocities, and in the
\textbf{relativistic} case, that includes the case of fast moving
particles with arbitrary high velocities, such as, for example,
particles of ionized gas in some medias at very high temperature
as : burning reactors, solar winds, nebular galaxies, etc...

1) In the non-relativistic case, the original result is
  due to Carleman, T. , in \cite{carleman} ;Di Blasio, G. in \cite{di blasio} Illner, R. , and
  Shinbrot, M. in \cite{illner}, prove a global result in the
  case of small data and without symmetric assumption ; an
  analogous result is unknown in the full relativistic case.

2) In the relativistic case, several authors proved local
  existence theorems, considering this equation alone , as
  Bichteler, K, in \cite{bichteler} ; Bancel, D. ,in \cite{bancel1} ,
  or coupling it to other fields equations as Bancel, D. , and
  Choquet-Bruhat, Y. , in \cite{bancel2} . Glassey, R. T. , and Strauss, W. ,
  obtained a global result in \cite{glassey1} , in the case of data
  near to that of an equilibrium solution with non-zero
  density.

  Now it would be interesting to extend to the
  relativistic case, the global existence theorem in the case of
  small initial data and even for large initial data. Mucha, P. B. , studied in
\cite{mucha1} and \cite{mucha2} , the relativistic Boltzmann
equation coupled to Einstein's equations, for a flat
Robertson-Walker space-time.

In this paper, we begin by giving, following Glassey, R. T. , in
\cite{glassey2}, the  formulation of the relativistic Boltzmann
equation, on a flat Robertson Walker space-time background, with a
given cosmological expansion factor.We consider the homogeneous
case which means that the distribution function depends only on
the time and the momentum of the particles. Such cases are very
useful for instance in cosmology. Next, we built a suitable
functional framework appropriate to the coupling with Einstein's
equations, and we construct a sequence of operators , to
approximate the collision operator, and that provide a sequence of
real-valued functions that converges to a global solution of the
relativistic Boltzmann equation. Moreover, we establish for this
sequence, a very simple a priori estimation, that leads to a
considerable simplification of the method followed in
\cite{mucha1} . We prove this way, under the only assumption that
the cosmological expansion factor is bounded away from zero, that
the initial valued problem for the relativistic Boltzmann equation
on the flat Robertson-Walker space-time , has an unique global
solution, provided that the initial data is sufficiently small,
and that the solution admits a very simple estimation by the
initial data. This paper extends the results of
\cite{noutchegueme} to the case of the curved Robertson-Walker
background space-time.

The paper is organized as follows : In section 2, we specify the
notations, we define the function spaces and we introduce the
relativistic Boltzmann equation and the collision operator. We end
this section by a sketch of the strategy adopted. In section 3, we
give the properties of the collision operator, some preliminary
results, and we construct the approximating operators. Section 4
is devoted to the global existence theorem.
\section{Notations, function spaces and the relativistic Boltzmann equation}
\subsection{Notations and function spaces}

  A Greek index varies from $0$ to $3$ and a Latin index from $1$ to
$3$. We consider as back-ground space-time, the flat
Robertson-Walker space-time denoted $(\mathbb{R}^{4}, g)$, where,
for $x = ( x^{\alpha} ) = (x^{o}, x^{i}) \in \mathbb{R}^{4}$,
$x^{o} = t$  denotes the time and $\bar{x} = (x^{i})$ the space ;
$g$ stands for the metric tensor with signature $(-, +, +, +)$
that can be written :
\begin{equation} \label{eq:2.1}
g = -dt^{2} + a^{2}(t)[(dx^{1})^{2} + (dx^{2})^{2} + (dx^{3})^{2}]
\end{equation}
in which, $a(t)$ is a given non-negative regular function of $t$ ,
called the cosmological expansion factor. We suppose $a(t)$
bounded from below, i.e :
\begin{equation} \label{eq:2.2}
a(t) > C_{o}
\end{equation}
where $C_{o} > 0$ is a given constant.

We consider the collisional evolution of a given kind of uncharged
particles in the time oriented curved space-time $(\mathbb{R}^{4},
g)$ . The distribution function of the particles we denote by $f$,
is a non-negative real-valued function of both the position
$(x^{\alpha})$ and the $4$-momentum $p = ( p^{\alpha} )$ of the
particles, and that coordinalize the tangent bundle
$T(\mathbb{R}^{4})$ i.e :
\begin{equation*}
f : T(\mathbb{R}^{4}) \simeq \mathbb{R}^{4} \times \mathbb{R}^{4}
\rightarrow \mathbb{R}^{+}, \quad (x^{\alpha}, p^{\alpha}) \mapsto
f(x^{\alpha}, p^{\alpha}) \in \mathbb{R}^{+}.
\end{equation*}
For $p = (p^{o}, p^{i}) = (p^{o}, \bar{p})$, and $q = (q^{o},
\bar{q})$ we set :
\begin{equation} \label{eq:2.3}
\bar{p} . \bar{q} = a^{2}(t)(p^{1}q^{1} + p^{2}q^{2} +p^{3}q^{3});
\quad \mid \bar{p} \mid^{2} =a^{2}(t)((p^{1})^{2} + (p^{2})^{2} +
(p^{3})^{2})
\end{equation}

We consider massive particles with a rest mass that can be
rescaled to $m = 1$. The particles are required  to move on the
future sheet of the mass-shell whose equation is $g(p, p) = -1$ ,
or equivalently, using (\ref{eq:2.1}) and (\ref{eq:2.3}) :
\begin{equation}\label{eq:2.4}
p^{o} = \sqrt{1 + \mid\bar{p}\mid^{2}}
\end{equation}

Hence, $f$ is in fact defined on the subbundle of
$T(\mathbb{R}^{4})$ defined by $(2.4)$ and whose local coordinates
are $x^{\alpha}$ and $p^{i}$. Now we consider the homogeneous case
for which $f$ depends only on the time $x^{o} = t$ and $\bar{p} =
(p^{i})$. The framework we will refer to is
$L^{1}(\mathbb{R}^{3})$ whose norm is denoted
$\parallel.\parallel$ ; \quad we set, for $r\in\mathbb{R}$,
$r>0$\, :
\begin{equation*}
X_{r} = \{ f \in L^{1}(\mathbb{R}^{3}), f \geq 0 \quad a. e ,
\parallel f \parallel \leq r \}
\end{equation*}
Endowed with the distance induced by the norm $\parallel .
\parallel$, $X_{r}$ is a complete and connected metric space. Let
$I$ be a real interval; we denote by $C[I;L^{1}(\mathbb{R}^{3})]$
the Banach space of continuous  and bounded functions $g:
I\rightarrow L^{1}(\mathbb{R}^{3})$ , endowed with   the norm :
$\parallel\mid g
\parallel\mid = \underset{t \in I}{Sup} \parallel g(t)
\parallel$. We set :
\begin{equation*}
C[I; X_{r}] = \{ g \in C[ I ; L^{1}(\mathbb{R}^{3}) ]\,,\, g(t)
\in X_{r} \,, \forall t \in I \, \, \}
\end{equation*}
Endowed with the distance induced by the norm $\parallel\mid
.\parallel\mid$ ;  $C[I; X_{r}]$ is a complete metric space.
\subsection{The relativistic Boltzmann equation}
The relativistic Boltzmann equation for uncharged particles in the
homogeneous case we consider , on the Robertson-Walker curved
space-time $(\mathbb{R}^{4}, g)$ can be written :
\begin{equation} \label{eq:2.5}
\frac{\partial  f}{\partial t} - 2\,\frac{\dot{a}}{a}\, \sum_{i =
1}^{3}p^{i} \frac{\partial f}{\partial p^{i}} = \frac{1}{p^{o}}
Q(f, f)
\end{equation}
with $\dot{a} = \frac{da}{dt}$, and where $Q$ is the collision
operator we now introduce. In the case of binary and elastic
collisions, $Q$ is defined as follows, $p$ and $q$ standing for
the momenta of two particles before their collision, $p'$ and $q'$
standing for their momenta after the collision, and where $f$ and
$g$ are two functions on $\mathbb{R}^{3}$ :

1)
\begin{equation}\label{eq:2.6}
\quad Q(f, g) = Q^{+}(f, g) - Q^{-}(f, g)
\end{equation}
with :

2)
\begin{equation}\label{eq:2.7}
\quad Q^{+}(f, g) = \int_{\mathbb{R}^{3}}
\frac{a^{3}(t)d\bar{q}}{q^{0}} \int_{S^{2}}f(\bar{p}')g(\bar{q}'
)A(t, \bar{p}, \bar{q}, \bar{p}', \bar{q}') d\omega
\end{equation}

3)
\begin{equation}\label{eq:2.8}
\quad Q^{-}(f,g) =\int_{\mathbb{R}^{3}}
\frac{a^{3}(t)d\bar{q}}{q^{0}}\int_{S^{2}}f(\bar{p})g(\bar{q})A(t,
\bar{p}, \bar{q}, \bar{p}', \bar{q}')d\omega
\end{equation}
in which;

 4)  $S^{2}$ is the unit sphere of $\mathbb{R}^{3}$ whose
area element is denoted $d\omega$;

5) $A$ is a non-negative real valued function of the indicated
arguments, called the collision kernel or the cross section of the
collisions and for which we require  the boundedness and the
symmetry assumptions :
\begin{equation}\label{eq:2.9}
0\leq A(t,\, \bar{p},\, \bar{q},\, \bar{p}',\, \bar{q}') \leq
C_{1}
\end{equation}
\begin{equation}\label{eq:2.10}
A(t, \bar{p},\, \bar{q},\, \bar{p}',\, \bar{q}') =A(t, \,
\bar{p}',\, \bar{q}', \, \bar{p},\, \bar{q})
\end{equation}
where $C_{1}$ is a given constant.

6) As consequences of the conservation law $p + q = p' + q'$, that
gives :
\begin{equation}\label{eq:2.11}
p^{o} +  q^{o} = p'^{o} +  q'^{o}\,; \quad  \bar{p} + \bar{q} =
\bar{p}' + \bar{q}'
\end{equation}

i) we have, using (\ref{eq:2.4}) and the first equation
(\ref{eq:2.11}) :
\begin{equation}\label{eq:2.12}
\sqrt{1 + \mid \bar{p}\mid^{2}} + \sqrt{1 + \mid \bar{q}\mid^{2}}
= \sqrt{1 + \mid \bar{p}'\mid^{2}} + \sqrt{1 +
\mid\bar{q}'\mid^{2}}
\end{equation}
which expresses the conservation of the quantity :
\begin{equation}\label{eq:2.13}
e =  \sqrt{1 + \mid\bar{p}\mid^{2}} + \sqrt{1 +
\mid\bar{q}\mid^{2}}
\end{equation}
called the energy of the unit rest mass particles ;

ii) we set, to express the second equation (\ref{eq:2.11}) and
following Glassey R.T, in \cite{glassey2} :
\begin{equation}\label{eq:2.14}
\begin{cases}
\bar{p}' = \bar{p} + b(\bar{p}, \bar{q}, \omega )\omega\\
\bar{q}' =\bar{p} - b(\bar{p}, \bar{q}, \omega)\omega \, ; \qquad
\omega \in S^{2}
\end{cases}
\end{equation}

in which $b(\bar{p}, \bar{q}, \omega)$ is a real-valued function.
Then, (\ref{eq:2.12}) and (\ref{eq:2.14}) lead to a quadratic
equation in $b$ that solves to give the only non trivial solution:
\begin{equation} \label{eq:2.15}
b(\bar{p}, \bar{q}, \omega) = \frac{2p^{o}q^{o} e\,
\omega.(\hat{\bar{q}} - \hat{\bar{p}})}{e^{2} - (\omega.(\bar{p} +
\bar{q}))^{2}}
\end{equation}
with $\hat{\bar{p}} = \frac{\bar{p}}{p^{o}}$, $e$ defined by
(\ref{eq:2.13}) and the dot product by (\ref{eq:2.3}). The
Jacobian of the change of variables $(\bar{p}, \bar{q}) \mapsto
(\bar{p}', \bar{q}')$ defined by (\ref{eq:2.14}) is given, using
the usual properties of determinants, by :
\begin{equation} \label{eq:2.16}
\frac{\partial(\bar{p}', \bar{q}')}{\partial(\bar{p}, \bar{q})} =
-\frac{p'^{o}q'^{o}}{p^{o}q^{o}}
\end{equation}

Notice that formulae (\ref{eq:2.15}) and (\ref{eq:2.16}) are
generalizations to the Robertson-Walker space-time, of analogous
formulae established in Glassey, R. T. , \cite{glassey2}, in the
case of the Minkowski space-time whose metric corresponds to the
particular case of (\ref{eq:2.1}) where $a(t) = 1$. It appears
clearly, using (\ref{eq:2.4}) and (\ref{eq:2.14}) that the
functions in the integrals (\ref{eq:2.7}) and (\ref{eq:2.8})
depend only on $\bar{p},  \bar{q},  \omega$ and that these
integrals give two functions $Q^{+}(f, g)$, $Q^{-}(f, g)$ of the
single variable $\bar{p}$. In practice, we will consider functions
$f$ on $\mathbb{R}^{4}$ that induce for $t\in \mathbb{R}$,
functions $f(t)$ on $\mathbb{R}^{3}$, defined by : $f(t)(\bar{p})
= f(t, \,\bar{p})$

Now, solving the Boltzmann equation (\ref{eq:2.5}) that is a first
order $p. d. e$, is equivalent to solving the associated
characteristic system that writes, taking $t$ as parameter:
\begin{equation}\label{eq:2.17}
\frac{dp^{i}}{dt} = -2\,\,\frac{\dot{a}}{a};\quad \frac{df}{dt} =
\frac{1}{p^{o}}Q(f, f)
\end{equation}

 We study the initial valued problem for the system (\ref{eq:2.17})
with initial data:
\begin{equation}\label{eq:2.18}
p^{i}(0) = y^{i}\,; \quad f(0) = f_{0}
\end{equation}

In (\ref{eq:2.17}), the equations in $p^{i}$ integrate directly to
give, setting  $y = (y^{i}) \in \mathbb{R}^{3}$;
\begin{equation}\label{eq:2.19}
\bar{p}(t, y) = \frac{a^{2}(0)}{a^{2}(t)}y
\end{equation}
whereas the initial valued problem  for $f$ is equivalent to the
integral equation in $f$:
\begin{equation}\label{eq:2.20}
f(t, y) = f_{o}(y) + \int_{0}^{t}\frac{1}{p^{o}}Q(f, f)(s, y)ds
\end{equation}
 Finally, solving the initial valued problem for the
relativistic Boltzmann equation (\ref{eq:2.5})with initial data
$f_{o}$, is equivalent to solving the integral equation
(\ref{eq:2.20}). The strategy we adopt to solve (\ref{eq:2.20})
will consist of :

a) Constructing an approximating operator $Q_{n}$ , with suitable
properties, that will converge point-wise in
$L^{1}(\mathbb{R}^{3})$ to $\frac{1}{p^{o}}Q$.

b) Solving by usual methods, the following approximating integral
equation:
\begin{equation} \label{eq:2.21}
f(t, y) = f_{o}(y) + \int_{0}^{t}Q_{n}(f,f)(s, y)ds
\end{equation}
obtained by replacing in (\ref{eq:2.20}),\, $\frac{1}{p^{o}}Q$ by
$Q_{n}$, and to obtain a global solution $f_{n}$ that will
converge to a global solution $f$ of (\ref{eq:2.20}), in a
suitable function space, that will be, given the general
framework, a space $C[0, + \infty; X_{r}]$ for a convenient real
number $r> 0$. Notice that, by (\ref{eq:2.19}), in
$\mathbb{R}^{3}$, the volume elements $d\bar{p}$,  $dy$ and the
corresponding norms $\parallel . \parallel_{\bar{p}}$ and
$\parallel . \parallel_{y}$ we adopt to be $\parallel .
\parallel$,  are linked by \,:
\begin{equation}\label{eq:2.22}
d\bar{p} = \frac{a^{6}(0)}{a^{6}(t)}dy ;\quad \parallel .
\parallel_{\bar{p}} \, = \,\frac{a^{6}(0)}{a^{6}(t)}\parallel . \parallel
\end{equation}

\section{Preliminary results and approximating operators.}
In this section, we give properties of the collision operator $Q$,
we construct and we give the properties of a sequence of
approximating operators $Q_{n}$ to $\frac{1}{p^{o}}Q$ .
\begin{proposition} \label{P:3.1}
If $f, g \in L^{1}(\mathbb{R}^{3})$, then $Q^{+}(f, g), Q^{-}(f,
g) \in L^{1}(\mathbb{R}^{3})$ and :
\begin{equation}\label{eq:3.1}
\parallel \frac{1}{p^{o}}Q^{+}(f, g)\parallel \leq C\parallel f \parallel
\parallel g \parallel
\end{equation}
\begin{equation}\label{eq:3.2}
\parallel \frac{1}{p^{o}}Q^{-}(f, g)\parallel \leq C\parallel f \parallel
\parallel g \parallel
\end{equation}
\begin{equation}\label{eq:3.3}
\parallel \frac{1}{p^{o}}Q^{+}(f, f) - \frac{1}{p^{o}}Q^{+}(g,
g)\parallel \leq C(\parallel f \parallel + \parallel g
\parallel)\parallel f - g \parallel
\end{equation}
\begin{equation}\label{eq:3.4}
\parallel \frac{1}{p^{o}}Q^{-}(f, f) - \frac{1}{p^{o}}Q^{-}(g,
g)\parallel \leq C(\parallel f \parallel + \parallel g
\parallel)\parallel f - g \parallel
\end{equation}
\begin{equation} \label{eq:3.5}
\parallel \frac{1}{p^{o}}Q(f, f) - \frac{1}{p^{o}}Q(g,
g)\parallel \leq C(\parallel f \parallel + \parallel\ g \parallel)
\parallel f - g \parallel
\end{equation}
where $C=\frac{8 \pi C_{1}a^{6}(0)}{C_{0}^{3}}$
\end{proposition}
\textbf{Proof:} 1) the expression (\ref{eq:2.7}) of $Q^{+}(f, g)$
gives, using the upper bound (\ref{eq:2.9}) of $A$ :
\begin{equation*}
\parallel \frac{1}{p^{o}} Q^{+}(f, g) \parallel_{\bar{p}} \leq C_{1}
a^{3}(t)\int_{\mathbb{R}^{3}}\int_{\mathbb{R}^{3}}
\frac{d\bar{p}d\bar{q}}{p^{o}q^{o}}\int_{S^{2}} \mid f(\bar{p}')
\mid \mid f(\bar{q}')\mid d\omega \qquad (a)
\end{equation*}
Now the change of variable $(\bar{p},\bar{q})\longmapsto(\bar{p}',
\bar{q}')$ defined by (\ref{eq:2.14}) and its Jacobian
(\ref{eq:2.16}) give: $d\bar{p}d\bar{q} =
\frac{p^{o}q^{o}}{p'^{o}q'^{o}}d\bar{p'}d\bar{q'}$ ;  then we
deduce from $(a)$, using (\ref{eq:2.4}) that gives
$\frac{1}{p'^{o}q'^{o}} \leq 1$ :
\begin{equation*}
\parallel\frac{1}{p^{o}} Q^{+}( f, g )\parallel_{\bar{p}} \leq C_{1} a^{3}(t)\int_{\mathbb{R}^{3}}\mid
f(\bar{p}')\mid d\bar{p}'\int_{\mathbb{R}^{3}}\mid g(\bar{q}')\mid
d\bar{q}' \int_{S^{2}} d\omega = 4 \pi C_{1} a^{3}(t)
\parallel f \parallel_{\bar{p}}\parallel g \parallel_{\bar{p}}
\end{equation*}
(\ref{eq:3.1}) then follows from the second relation
(\ref{eq:2.22}) and  (\ref{eq:2.2}) . $\Box$

2) The expression (\ref{eq:2.8}) for $Q^{-}(f, g)$ gives, using
(\ref{eq:2.9}) and (\ref{eq:2.4}) that gives $\frac{1}{p^{o}q^{o}}
\leq 1$ ;
\begin{equation*}
\parallel\frac{1}{p^{o}} Q^{-}( f, g )\parallel_{\bar{p}} \leq C_{1} a^{3}(t)\int_{\mathbb{R}^{3}}\mid
f(\bar{p})\mid d\bar{p}\int_{\mathbb{R}^{3}}\mid g(\bar{q})\mid
d\bar{q} \int_{S^{2}} d\omega = 4 \pi C_{1} a^{3}(t)
\parallel f \parallel_{\bar{p}}\parallel g \parallel_{\bar{p}}
\end{equation*}
and (\ref{eq:3.2}) follows from the second relation
(\ref{eq:2.22}) and (\ref{eq:2.2}) $\Box$

3) Inequalities (\ref{eq:3.3}), (\ref{eq:3.4}), (\ref{eq:3.5})are
direct consequences of (\ref{eq:3.1}), (\ref{eq:3.2}), \\ $Q =
Q^{+} - Q^{-}$, and the fact that, the expressions (\ref{eq:2.7})
and (\ref{eq:2.8}) show that $Q^{+}$ and  $Q^{-}$ are bilinear
operators, and so we can write, $P$ standing for
$\frac{1}{p^{o}}Q^{+}$ or $\frac{1}{p^{o}}Q^{-}$ : $P(f, f) - P(g,
g) = P(f, f - g) + P(f-g, g)$. This completes the prove of
proposition (\ref{eq:3.1}) $\Box$

We now state and prove the following result on which relies the
construction of the approximating operators ; $C$ denotes the
constant defined in proposition(\ref{eq:3.1}).

\begin{proposition} \label{P:3.2}
Let $r$ be an arbitrary strictly positive real number. Then, there
exists an integer $n_{o}(r) > 0$ such that, for every $n >
n_{o}(r)$ and for every $v \in X_{r}$, the equation :
\begin{equation}\label{eq:3.6}
\sqrt{n} u \,- \,\frac{1}{p^{o}} Q(u, u) \,= \, v
\end{equation}
has a unique solution $u_{n} \in X_{r}$
\end{proposition}

\textbf{Proof} Let $ r > 0$ be given and let $ v \in X_{r}$. We
can write (3.6), using $Q = Q^{+} - Q^{-}$ :
\begin{equation*}
\sqrt{n} u \,- \,\frac{1}{p^{o}} Q^{+}(u, u) + \,\frac{1}{p^{o}}
Q^{-}(u, u) \,= \, v
\end{equation*}
Notice that one deduces from the expression (2.8) of $Q^{-}$,
since $\bar{p}$ is not a variable for that integral, that
 $Q^{-}(f, g) = f Q^{-}(1, g)$. Hence the above equation can be
written, using $Q^{-}(u, u) = u Q^{-}(1, u)$, for $u \geq 0$ a.e.
\begin{equation*}
  u \, = \, F_{n}(u) \tag{a}
\end{equation*}
where
\begin{equation*}
F_{n}(h) = \frac{v + \frac{1}{p^{o}} Q^{+}(h, h)}{\sqrt{n} +
\frac{1}{p^{o}} Q^{-}(1, h)} \tag{b}
\end{equation*}
(a) shows that a solution $u$ of (3.6) can be considered as a
fixed point of the map $h \longmapsto F_{n}(h)$ defined by (b).
Let us shows that  $F_{n}$ is a contracting map from $X_{r}$ to
$X_{r}$. We will then solve the problem by applying the fixed
point theorem .

i)We deduce from (3.1) that $F_{n} : X_{r} \longmapsto
L^{1}(\mathbb{R}^{3})$ , and that since $v , h \in X_{r}$ :
\begin{equation*}
\parallel F_{n}(h)\parallel \, \leq \, \frac{\parallel v
\parallel}{\sqrt{n}}\, + \, \frac{C\parallel h \parallel \,\parallel h
\parallel}{\sqrt{n}}\, \leq \,\frac{r(1+Cr)}{\sqrt{n}} \tag{c}
\end{equation*}
ii) Let $g,\, h \in X_{r}$ . (b) gives :
\begin{equation*}
F_{n}(h) - F_{n}(g) = \frac{v + \frac{1}{p^{o}} Q^{+}(h,
h)}{\sqrt{n} + \frac{1}{p^{o}} Q^{-}(1, h)} - \frac{v +
\frac{1}{p^{o}} Q^{+}(g, g)}{\sqrt{n} + \frac{1}{p^{o} }Q^{-}(1,
g)} \tag{d}
\end{equation*}
Since $g \geq 0$, $\,h \geq 0$ a.e., we have , taken $n \geq 0$:
\begin{equation*}
  [\sqrt{n} + \frac{1}{p^{o}} Q^{-}(1, h)][\sqrt{n} + \frac{1}{p^{o}} Q^{-}(1, g)]
\, \geq \, n \geq \sqrt{n}.
\end{equation*}
We then deduce from (d), using once more $Q^{-}(f, g) = fQ^{-} (1,
g )$, that :
\begin{align*}
\sqrt{n} \parallel F_{n}(h) - F_{n}(g) \parallel &\leq
\parallel \frac{1}{p^{o}} Q^{+}(h, h) - \frac{1}{p^{o}} Q^{+}(g, g) \parallel
+ \parallel \frac{1}{p^{o}} Q^{-}(v, g) - \frac{1}{p^{o}} Q^{-}(v,
h) \parallel\\
& \qquad  + \parallel \frac{1}{p^{o}} Q^{-}[\frac{1}{p^{o}}
Q^{+}(h, h), g] - \frac{1}{p^{o}} Q^{-}[\frac{1}{p^{o}} Q^{+}(g,
g), h] \parallel \qquad  (e)
\end{align*}

We can write, using the bilinearity of $Q^{-}$ :
\begin{equation*}
\frac{1}{p^{o}}\, Q^{-}(v, g) - \frac{1}{p^{o}}\, Q^{-}(v, h) =
\frac{1}{p^{o}}\, Q^{-}(v, g - h)
\end{equation*}
So, in (e) we apply (3.3) to the first term, (3.2) to the second
term to obtain, using $\parallel g \parallel \leq r$, $\parallel h
\parallel \leq r$, $\parallel v \parallel \leq r$ :
\begin{equation*}
\parallel \frac{1}{p^{o}}\, Q^{+}(h, h) - \frac{1}{p^{o}}\, Q^{+}(g, g) \parallel
 + \parallel \frac{1}{p^{o}}\, Q^{-}(v, g) + \frac{1}{p^{o}}\,Q^{-}(v, h)
\parallel \leq 3Cr\parallel h - g \tag{f}
\end{equation*}
Concerning the last term in (e), we can write, using the
bilinearity of $Q^{-}$ :
\begin{align*}
\frac{1}{p^{o}}\, Q^{-}\left[\frac{1}{p^{o}}\,Q^{+}(h, h), g
\right] - \frac{1}{p^{o}}\,Q^{-}\left[\frac{1}{p^{o}}\, Q^{+}(g,
g), h \right] &= \frac{1}{p^{o}}\, Q^{-}\left[\frac{1}{p^{o}}\,
Q^{+}(h, h),
g -h \right]\\
& \qquad  +  \frac{1}{p^{o}}\, Q^{-}\left[\frac{1}{p^{o}}\,
Q^{+}(h, h) - \frac{1}{p^{o}}\,Q^{+}(g, g), h \right]
\end{align*}
We then apply to the right hand side of the above equality :

1) For the first term, property (3.2) of $Q^{-}$, followed by
property (3.1) of $Q^{+}$

2) For the second term, property (3.2) of $Q^{-}$, followed by
property (3.3) of $Q^{+}$ to obtain, using $\parallel g \parallel
\leq r$, $\parallel h \parallel \leq r$ :
\begin{equation*}
\parallel \frac{1}{p^{o}} Q^{-}\left[\frac{1}{p^{o}} Q^{+}(h, h), g \right]
- \frac{1}{p^{o}} Q^{-}\left[\frac{1}{p^{o}} Q^{+}(g, g), h
\right]
\parallel \, \leq \, 3C^{2}r^{2}\parallel h - g \parallel  \tag{h}
\end{equation*}
thus (e) gives, using (f) and (h) :
\begin{equation*}
 \parallel F_{n}(h) - F_{n}(g) \parallel \leq \frac{3Cr(1 + Cr)}{\sqrt{n}}
\parallel h - g \parallel \tag{i}
\end{equation*}

It is then easy , using (c) and (i), to choose an interger
$n_{o}(r)$ such that, for $n \geq n_{o}(r)$ , we have:
\begin{equation}\label{eq:3.7}
\frac{1 + Cr}{\sqrt{n}} \leq 1 \quad and \quad \frac{3Cr(1 +
Cr)}{\sqrt{n}} \leq \frac{1}{2}
\end{equation}
This shows , using once more (c) and (i) that, for $n \geq
n_{0}(r)$, $F_{n}$ is a contracting map of the complete metric
space $X_{r}$ into itself. By the fixed point theorem, $F_{n}$ as
a unique fixed point $u_{n} \in X_{r}$ that is a unique solution
of (3.6).\\
 In all the following $n_{o}(r)$ stands for the
integer defined above.\\
 The above result leads us to the following definition :
\begin{definition}\label{d:3.1}
 Let $n \in
\mathbb{N}$, $n \geq n_{o}(r)$ be given.

1) Define the operator $R(n, Q) : X_{r} \longrightarrow
X_{r},\quad u \longmapsto R(n, Q)u$ as follows : for $u \in
X_{r}$, $R(n, Q)u$ is the unique element of $X_{r}$ such that :
\begin{equation}\label{eq:3.8}
 \sqrt{n}R(n, Q)u - \frac{1}{p^{o}} Q\left[R(n, Q)u, R(n, Q)u\right] = u
\end{equation}

2) Define the operator $Q_{n}$ on $X_{r}$ by :
\begin{equation}\label{eq:3.9}
  Q_{n}(u, u) = n\sqrt{n}R(n, Q)u - nu
\end{equation}
\end{definition}
We give the properties of the operators $R(n, Q)$ and $Q_{n}$.
\begin{proposition}\label{p:3.3}:
Let $r \in \mathbb{R}^{*}_{+}$, $n, m \in \mathbb{N}$ $m \geq
n_{o}(r)$, $n \geq n_{o}(r)$, be given.Then we have, for every $u,
v \in X_{r}$
\begin{eqnarray}
\parallel \sqrt{n} R(n, Q)u \parallel &  = & \parallel u \parallel \\
\parallel \sqrt{n} R(n, Q)u \,- \, u \parallel & \leq & \frac{K}{n} \\
\parallel R(n, Q)u \, - \,R(n, Q)v \parallel &  \leq &
\frac{2}{\sqrt{n}} \parallel u - v \parallel \\
Q_{n}(u, u) & = & \frac{1}{p^{o}} Q\left[ \sqrt{n} R(n, Q)u,
\sqrt{n} R(n, Q)u
\right] \\
\parallel Q_{n}(u, u) - Q_{n}(v, v) \parallel & \leq & K\parallel u - v \parallel
\\
\parallel Q_{n}(u, u) - \frac{1}{p^{o}} Q(v, v) \parallel & \leq &
\frac{K}{n} +  K\parallel u - v \parallel \\
\parallel Q_{n}(u, u) - Q_{m}(v, v) \parallel & \leq &
K\parallel u - v \parallel + \frac{K}{n} + \frac{K}{m}
\end{eqnarray}
Here $K = K(r)$ is a continuous function of $r$.
\end{proposition}
\textbf{Proof}. 1) (3.10) will be a consequence of the relation:
\begin{equation}\label{eq:3.17}
\int_{\mathbb{R}^{3}}\,\frac{1}{p^{o}} Q(f, g)\,d\bar{p} = 0,
\qquad \forall f, g \in L^{1} (\mathbb{R}^{3})
\end{equation}
we now establish. This is where we need assumption (2.10) on the
kernel $A$. We have, using (\ref{2.11}), definition (2.7) of
$Q^{+}$, the change of variables $(\bar{p}, \bar{q}) \longmapsto
(\bar{p'}\bar{q'})$ defined by (2.14) and its Jacobian (2.16), and
since $ p^{o} = \sqrt{1 + \mid \bar{p} \mid^{2}}$ :
\begin{equation*}
I_{o} = \int_{\mathbb{R}^{3}}\,\frac{1}{p^{o}} Q^{+}(f,
g)\,d\bar{p} = a^{3}(t)\int_{\mathbb{R}^{3}}\,d\bar{p'}
\int_{\mathbb{R}^{3}}\,d\bar{q'}\int_{S^{2}}\,\frac{f(\bar{p'})
g(\bar{q'}) S(\bar{p'}, \bar{q'}, \bar{p}, \bar{q}) }{\sqrt{1 +
\mid \bar{p'} \mid^{2}}\,\sqrt{1 + \mid \bar{q'}
\mid^{2}}}\,d\omega
\end{equation*}
On the other hand, we have, using definition (2.8) of $Q^{-}$ and
$ p^{o} = \sqrt{1 + \mid \bar{p} \mid^{2}}$ :
\begin{equation*}
J_{o} = \int_{\mathbb{R}^{3}}\,\frac{1}{p^{o}} Q^{-}(f,
g)\,d\bar{p} = a^{3}(t)\int_{\mathbb{R}^{3}}\,d\bar{p}
\int_{\mathbb{R}^{3}}\,d\bar{q}\int_{S^{2}}\,\frac{f(\bar{p})
g(\bar{q}) S(\bar{p}, \bar{q}, \bar{p'}, \bar{q'}) }{\sqrt{1 +
\mid \bar{p} \mid^{2}}\,\sqrt{1 + \mid \bar{q} \mid^{2}}}\,d\omega
\end{equation*}
It then appears that $I_{o} = J_{o}$ and (\ref{eq:3.17}) follows
from $Q = Q^{+} - Q^{-}$. Then, to obtain (3.10), we integrate
(3.8) over $\mathbb{R}^{3}$ with respect to $\bar{p}$, use (3.17),
and conclude by (2.22) that gives the equivalence between
$\parallel . \parallel_ {\bar{p}}$ and
$\parallel . \parallel$ $\Box$ \\
2) The definition (3.8) of $R(n, Q)$ gives, using property (3.5)
of $Q$ with \\$f =  R(n, Q)u$ and $g = 0$ :
\begin{equation*}
  \parallel \sqrt{n} R(n, Q)u - u \parallel = \parallel\frac{1}{p^{o}}
Q\left[R(n, Q)u, R(n, Q)u\right]\parallel \leq C
\parallel R(n, Q)u \parallel^{2}
\end{equation*}
and (3.11) follows from (3.10) that gives $\parallel R(n,
Q)u\parallel \leq \frac{\parallel u \parallel}{\sqrt{n}} \leq
\frac{r}{\sqrt{n}}$.\\
 3) Subtracting the relations (3.8) written for $u$ and $v$ gives:
\begin{equation*}
R(n, Q)u - R(n, Q)v = \frac{u - v}{\sqrt{n}} +
\frac{1}{\sqrt{n}p^{o}} Q\left[R(n, Q)u, R(n, Q)u\right] -
\frac{1}{\sqrt{n}p^{o}} Q\left[R(n, Q)v, R(n, Q)v\right]
\end{equation*}
which gives, using property (3.5) of $Q$ and once more (3.10):
\begin{equation*}
\parallel R(n, Q)u - R(n, Q)v \parallel \leq \frac{\parallel u - v \parallel}
{\sqrt{n}} + \frac{2Cr}{\sqrt{n}} \parallel R(n, Q)u - R(n, Q)v
\parallel
\end{equation*}
(3.12) then follows from $n \geq n_{o}(r)$ and (3.7) that implies
$\frac{2Cr}{\sqrt{n}} \leq \frac{1}{2}$.\\
4) The expression (3.13) of $Q_{n}$ is obtained by multiplying
(3.8) by $n$, using its definition (3.9) and the bilinearity of
$Q$.\\
5) Property (3.14) is obtained by using (3.13), property (3.5) of
$Q$, followed by properties (3.10) and (3.12) of $R(n, Q)$. \\
6) Notice that by (3.10), $\sqrt{n} R(n, Q)u \in X_{r}$ since $u
\in X_{r}$. Then, the expression (3.13) of $Q_{n}$ and property
(3.5) of $Q$ give :
\begin{equation*}
\parallel Q_{n}(u, u) - \frac{1}{p^{o}} Q(v, v) \parallel \leq 2Cr
\parallel \sqrt{n} R(n, Q)u - v \parallel \leq
2Cr\left(\parallel \sqrt{n} R(n, Q)u - u\parallel +
\parallel u - v \parallel\right)
\end{equation*}
hence, (3.15) follows from (3.11). \\
7) The expression (3.13) of $Q_{n}$ gives:
\begin{equation*}
Q_{n}(u, u) - Q_{m}(v, v) =  \frac{1}{p^{o}} Q\left[ \sqrt{n} R(n,
Q)u, \sqrt{n} R(n, Q)u \right] - \frac{1}{p^{o}} Q\left[ \sqrt{m}
R(m, Q)v, \sqrt{m} R(m, Q)v
\right] \\
\end{equation*}
then, (3.16) follows from (3.5), (3.10) and (3.11), adding and
subtracting $u$ and $v$. $\Box$

\begin{remark}.
(3.15) shows by taking $u = v$, that $Q_{n}$ converges pointwise
in the $L^{1}(\mathbb{R}^{3})$-norm to the operator
$\frac{1}{p^{o}} Q$. From there, the qualification of \\
''approximating operators'' we give to the sequence $Q_{n}$.
\end{remark}

We now have all the tools we need to prove the global existence
theorem.

\section{The global existence theorem}
With the approximating operator $Q_{n}$ defined by (3.8), we will
first state and prove a global existence theorem for the
''approximating'' integral equation:
\begin{equation}\label{eq:4.1}
  f(t_{o} + t, y) = g_{o}(y) + \int_{t_{o}}^{t_{o} +
t}Q_{n}\left(f, f\right)(s, y)\,dy , \qquad t \geq 0,
\end{equation}
in which $t_{o}$ is an arbitrary positive real number, and $g_{o}$
is a given function that stands for the initial data at time
$t_{o}$; for practical reasons,we study (\ref{eq:4.1}) rather than
(\ref{eq:2.21}) that is the particular case when $t_{o} = 0$ and
$g_{o} = f_{o}$. Next, we prove, by a convenient choice of $t_{o}$
that the global solution $f_{n}$ of (\ref{eq:4.1}) converges in a
sense to specify, to a global solution $f$ of (\ref{eq:2.20}).
\begin{proposition}\label{p:4.1}
Let $r$ be an arbitrary strictly positive real number. Let $g_{o}
\in X_{r}$ and let $n$ be an integer such that $n \geq n_{o}(r)$.
Then, for every $t_{o} \in [0, +\infty[$, the integral equation
(\ref{eq:4.1}) has a unique solution $f_{n} \in C[t_{o}, +\infty;
X_{r}]$. Moreover, $f_{n}$ satisfies the inequality :
\begin{equation}\label{eq:4.2}
\mid\parallel f_{n} \mid\parallel \leq \parallel g_{o} \parallel
\end{equation}
\end{proposition}
\textbf{Proof}: \\
We give the proof in 2 steps. \\
\textbf{Step 1}:Local existence and estimation. \\
Consider equation (\ref{eq:4.1}) for $t \in [t_{o}, t_{o} +
\delta]$ where $\delta  >0$ is given. One verifies easily, using
(3.8) that gives : $ Q_{n}(f, f) + n f = n^{2}R(n, Q)f $  and \\$
\frac{d}{dt} [e^{nt}f(t_{o} + t, y)] = e^{nt}\left[ \frac{df}{dt}
+ nf \right](t_{o} + t, y) $, that (\ref{eq:4.1}) is equivalent to
:
\begin{equation}\label{eq:4.3}
 f(t_{o} + t, y) = e^{-nt}g_{o}(y) + \int_{0}^{
t} n\sqrt{n} e^{-n(t - s)}R(n, Q)f(t_{o} + s, y)\,dy
\end{equation}
We solve (\ref{eq:4.3}) by the fixed point theorem. Let us define
the operator $A$ over the complete metric space $C[t_{o}, t_{o} +
\delta; X_{r}]$ by :
\begin{equation}\label{eq:4.4}
Af(t_{o} + t, y) = e^{-nt}g_{o}(y) + \int_{0}^{ t}n\sqrt{n}
e^{-n(t - s)}R(n, Q)f(t_{o} + s, y)\,dy
\end{equation}
\begin{itemize}
  \item [i)] Let $f \in C[t_{o}, t_{o} + \delta; X_{r}]$;
(\ref{eq:4.4}) gives, since $\parallel g_{o}\parallel \leq r$ and
using (3.9) that gives, $\parallel \sqrt{n} R(n, Q)f(t_{o} +
s)\parallel = \parallel f(t_{o} + s)\parallel\, \leq r $, and for
every $t \in [0, \delta]$ :\begin{equation*}
\parallel Af(t_{o} + t) \parallel\, \leq \,e^{-nt}r + nre^{-nt}\int_{0}^{t}
e^{ns}\,ds = e^{-nt}r + re^{-nt}(e^{nt} - 1) = r
\end{equation*}
Hence $\mid\parallel Af \mid\parallel \leq r$, and this shows that
$A$ maps $C[t_{o}, t_{o} + \delta; X_{r}]$ into itself. \item[ii)]
Let $f, g \in C[t_{o}, t_{o} + \delta; X_{r}]$; $t \in [0,
\delta]$; (4.4)gives:
\begin{equation*}
 \left(Af - Ag\right)(t_{o} + t, y) = \int_{0}^{ t} n\sqrt{n} e^{-n(t -
s)}\left[R(n, Q)f - R(n, Q)g\right](t_{o} + s, y)\,dy
\end{equation*}
which gives, using property (3.12) of $R(n, Q)$ and $e^{-n(t - s)}
\leq 1$ :
\begin{equation*}
\mid\parallel Af - Ag \mid\parallel \leq 2n\delta\mid\parallel f -
g \mid\parallel.
\end{equation*}
 \end{itemize}
It then appears that $A$ is a contracting map in a metric space
$C[t_{o}, t_{o} + \delta; X_{r}]$ where \, $2\,n\,\delta \leq
\frac{1}{2}$, i.e. $\delta \in ]0, \frac{1}{4n}]$. Taking $\delta
= \frac{1}{4n} $, we conclude that $A$ has a unique fixed point
  $f_{n}^{o} \in C[t_{o}, t_{o} + \frac{1}{4n}; X_{r}]$, that is
the unique solution of (4.3) and hence, of (4.1). \\
Now we have, by taking $t = 0$ in (4.3), $f(t_{o}, y) = g_{o}(y)$.
The solution $f_{n}^{o}$ then satisfies:
\begin{equation*}
 f_{n}^{o}(t_{o} + t, y) = e^{-nt}f_{n}^{o}(t_{o}, y) + \int_{0}^{
t} n\sqrt{n} e^{-n(t - s)}R(n, Q)f_{n}^{o}(t_{o} + s, y)\,dy
\end{equation*}
which gives, multiplying by $e^{nt}$, using once more (3.10), and
for $t \in ]0, \frac{1}{4n}]$:
\begin{equation*}
 \parallel e^{nt}f_{n}^{o}(t_{o} + t) \parallel\, \leq \, \parallel f_{n}^{o} \parallel
 + n\int_{0}^{t}\,\parallel e^{ns}f_{n}^{o}(t_{o} + s) \parallel \,ds
\end{equation*}
this gives, by Gronwall Lemma: $e^{nt}\parallel f_{n}^{o}(t_{o} +
t )\parallel \leq e^{nt}\parallel f_{n}^{o}(t_{o})\parallel$ and
hence
\begin{equation}\label{eq:4.5}
 \mid\parallel f_{n}^{o} \mid\parallel\, \leq \, \parallel f_{n}^{o}(t_{o}) \parallel
\end{equation}

\textbf{Step 2:} Global existence; estimation and uniqueness. \\
Let $k \in \mathbb{N}$. By replacing in the integral equation
(4.1) $t_{o}$ by : \\ \mbox{$t_{o} + \frac{1}{4n}, t_{o} +
\frac{2}{4n}, \cdots, t_{o} + \frac{k}{4n}, \cdots$}, step 1 tells
us that on each interval \\ $I_{k} = \left[t_{o} + \frac{k}{4n},
t_{o} + \frac{k}{4n} + \frac{1}{4n}\right]$ whose length is
$\frac{1}{4n}$, the initial value problem for the corresponding
integral equation has a unique solution $f_{n}^{k} \in C[I_{k},
X_{r}]$, provided that the initial data we denote $f_{n}^{k}\left[
t_{o} + \frac{k}{4n} \right]$ is a \textbf{given} element of
$X_{r}$; $f_{n}^{k}$ then satisfies:
\begin{equation*}
\begin{cases}
f_{n}^{k}(t_{o} + \frac{k}{4n} + t, y) = f_{n}^{k}(t_{o} +
\frac{k}{4n}, y) + \int_{t_{o} + \frac{k}{4n}}^{t_{o} +
\frac{k}{4n} + t}\, Q_{n}\left( f_{n}^{k},
f_{n}^{k}\right)(s,y)\,ds \\
f_{n}^{k}\left(t_{o} + \frac{k}{4n}\right) \in X_{r};\quad k \in
\mathbb{N}; \quad t \in \left[ 0, \frac{1}{4n}\right]
\end{cases}\end{equation*}
and (4.5)implies:
\begin{equation*}
  \mid\parallel f_{n}^{k} \mid\parallel \,\leq \,\parallel f_{n}^{k}(t_{o} +
 \frac{k}{4n}) \parallel.
\end{equation*}
Now, $[0, +\infty[ = \underset{k \in \mathbb{N}}{\bigcup} \,\left[
t_{o} + \frac{k}{4n}, t_{o} + \frac{k}{4n} + t \right]$; Define :
\begin{equation*}
f_{n}^{k}(t_{o} + \frac{k}{4n}, y) = f_{n}^{k - 1}(t_{o} +
\frac{k}{4n}, y)\quad \text{if} \quad k \geq 1 \, \text{and} \quad
f_{n}^{o}(t_{o}) = g_{o}.
\end{equation*}
Then, the solution $f_{n}^{k - 1}$ and $f_{n}^{k}$ that are
defined on $\left[  t_{o} + \frac{k - 1}{4n},  t_{o} +
\frac{k}{4n} \right]$ and $\left[  t_{o} + \frac{k}{4n},  t_{o} +
\frac{k + 1}{4n} \right]$ overlap at $t =  t_{o} + \frac{k}{4n}$.\\
Define the function $f_{n}$ on $[0, +\infty[$ by
\begin{equation*}
 f_{n}(t) = f_{n}^{k}(t) \quad \text{if} \quad t \in \left[  t_{o} +
\frac{k}{4n},  t_{o} + \frac{k + 1}{4n} \right].
\end{equation*}
Then, a straightforward calculation shows, using the above
relations for\\ \mbox{$k, k - 1, k - 2, \cdots, 1,0,$} that
$f_{n}$ is a global solution of (4.1) on $[0, +\infty[$ satisfying
the estimation (4.2) and hence, $f_{n} \in C[t_{o}, +\infty;
X_{r}]$.\\
Now suppose that $f, g \in C[t_{o}, +\infty; X_{r}]$ are 2
solutions of (4.1), then:
\begin{equation*}
\left( f - g \right)(t_{o} + t, y) = \int_{o}^{t}\left[ Q_{n}(f,
f) - Q_{n}(g, g)\right](t_{o} + s, y)\,ds \quad t \geq 0.
\end{equation*}
This gives, using property (3.14) of $Q_{n}$:
\begin{equation*}
 \parallel \left(f - g\right)(t_{o} + t)\parallel \,\leq \,K\int_{0}^{t}
\parallel \left(f - g\right)(t_{o} + s)\parallel \,ds \quad t \geq 0
\end{equation*}
which implies, using Gronwall Lemma, $f = g$, and the uniqueness
follows. This ends the proof of proposition 4.1.$\Box$

 We now state and prove the global existence theorem.
\begin{theorem}\label{T:4.1}
Let $f_{o} \in L^{1}(\mathbb{R}^{3})$, $f_{o} \geq 0$, $a.e$; be
arbitrary given.  Then the Cauchy problem for the relativistic
Boltzmann equation on the Robertson-Walker space-time, with
initial data $f_{o}$, has a unique global solution $f \in C[0,
+\infty; L^{1}(\mathbb{R}^{3})]$, $f(t) \geq 0, a.e, \forall t \in
[0, +\infty[$; $f$ satisfies the estimation:
\begin{equation}\label{eq:4.6}
  \underset{t \in [0, +\infty[}{Sup}\parallel f(t)\parallel \,\leq
\, \parallel f_{o} \parallel
\end{equation}
\end{theorem}
\textbf{Proof}:
Fix a real number $r>\parallel f_{o} \parallel$. We give the proof in 2 steps. \\
 \textbf{\underline{Step 1}}: Local existence and estimation.

Let $t_{o} \geq 0$ and $g_{o} \in X_{r}$ be given. The proposition
4.1 gives for every integer $n \geq n_{o}(r)$, the existence of a
solution $f_{n} \in C[t_{o}, +\infty; X_{r}]$ of (4.1) that
satisfies the estimation (4.2). It is important to notice that
this solution depends on $n$. \\
Let $T > 0$ be given;  we have $f_{n} \in C[t_{o}, t_{o} + T;
X_{r}], \,\forall n \in \mathbb{N}, \, n \geq n_{o}(r)$. Let us
prove that the sequence $f_{n}$ converges in $ C[t_{o}, t_{o} + T;
X_{r}]$ to a solution of the integral equation :
\begin{equation}\label{eq:4.7}
f(t_{o} + t, y) = g_{o}(y) + \int_{t_{o}}^{t_{o} + t}
\frac{1}{p^{o}}Q\left(f, f\right)(s, y)\,dy , \qquad t \in[0, T]:
\end{equation}
Consider 2 integers $m, n \geq n_{o}(r)$ ; we deduce from (4.1),
that, $\forall t \in [0, T]$;
\begin{equation*}
 f_{n}(t_{o} + t, y) - f_{m}(t_{o} + t, y)=  \int_{t_{o}}^{t_{o} +
t}\left[Q_{n}\left(f_{n}, f_{n}\right) - Q_{m}\left(f_{m}, f_{m}
\right)\right](s, y)\,dy
\end{equation*}
this gives, using the property (3.16) of the operators $Q_{n}$,
$Q_{m}$ and Gronwall lemma:$
 \mid\parallel f_{n} - f_{m} \mid\parallel \,\leq \,\left(\frac{1}{n}
+ \frac{1}{m} \right)KTe^{KT}$. This proves that $f_{n}$ is a
Cauchy sequence in the complete metric space \\$ C[t_{o}, t_{o} +
T, X_{r}]$. Then, there exists $f \in C[t_{o}, t_{o} + T, X_{r}]$
such that:
\begin{equation}\label{eq:4.8}
 f_{n} \,\text{ converges to }\,f \,\text{ in }\, C[t_{o}, t_{o} + T,
X_{r}]
\end{equation}
The definition of $X_{r}$ implies that $f(t_{o} + t) \geq 0 $
a.e;\\
 Let us show that $f$ satisfies (4.7). The convergence (4.8)
implies that, $f_{n}(t_{o} + t)$ converges to $f(t_{o} + t),
\,\forall t \in [0, T]$; then $f_{n}(t_{o})$ converges to
$f(t_{o})$; but $f_{n}(t_{o}) = g_{o}$, then $f(t_{o}) = g_{o}$;
next we have, using property (3.14) of $Q_{n}$ and $Q$ :
\begin{equation*}
\parallel \int_{t_{o}}^{t_{o} +
t}\left[Q_{n}\left(f_{n}, f_{n}\right) - \frac{1}{p_{o}} Q\left(f,
f \right)\right](s, y)\,dy \parallel \,\leq \,\frac{KT}{n} +
KT\mid\parallel f_{n} - f \mid\parallel
\end{equation*}
which shows that
\begin{equation*}
\int_{t_{o}}^{t_{o} + t}Q_{n}\left( f_{n}, f_{n} \right)(s)\,ds\,
\text{ converges to }\,\int_{t_{o}}^{t_{o} + t}Q\left(f,
f\right)(s)\,ds \,\text{ in }\,L^{1}(\mathbb{R}^{3}).
\end{equation*}
Hence, since $f_{n}$ satisfies (4.1), $f$ satisfies (4.7),\,
$\forall t \in [0, T]$ and a.e. in $y \in \mathbb{R}^{3}$.
Finally, (4.2) and (4.8) imply:
\begin{equation}\label{eq:4.9}
  \mid\parallel f \mid\parallel \,\leq \,\parallel f(t_{o})\parallel
\end{equation}

\textbf{\underline{Step 2}}: Global existence, estimation and
uniqueness

Since $t_{o} \in [0, +\infty[$ is arbitrary and since $\forall n
\geq n_{o}(r)$, the solution $f_{n}$ of (4.1) is global on
$[t_{o}, +\infty[$, Step 1 tells us that, given $T > 0$, by taking
in the integral equation (4.7), $t_{o} = 0, t_{o} = T, t_{o} = 2T,
\cdots, t_{o} = kT, \cdots$ $k \in \mathbb{N}$, then, on each
interval $J_{k} = [kT, (k + 1)T]$, $k \in \mathbb{N}$, whose
length is $T$, the initial value problem for the corresponding
integral equation has a solution $f^{k} \in C[J_{k}, X_{r}]$,
provided that the initial data we denote $f^{k}(kT)$, is a given
element of $X_{r}$. \\
We then proceed exactly as in Step 2 of the proof of Proposition
4.1  by writing for $T > 0$ given, that $[0, +\infty[ =
\underset{k \in \mathbb{N}}{\bigcup}[kT, (k + 1)T]$, to overlap
the local solutions $f^{k}\in C[kT, (k + 1)T; X_{r}]$, by setting
$f^{k}(kT) = f^{k - 1}(kT)$ if $k \geq 1$ and $f^{o}(0) = f_{o}$,
and we obtain a global solution $f \in C[0, +\infty; X_{r}]$ of
the integral equation (2.20) defined by $f(t) = f^{k}(t)$, if $t
\in [kT, (k + 1)T]$. Next, (4.9) that gives: $\mid\parallel
f^{k}\mid\parallel \, \leq \,\parallel f^{k}(kT)\parallel$,
shows that $f$ satisfies the estimation (4.6). \\
Now if $f, g \in C[0, +\infty; L^{1}(\mathbb{R}^{3})]$ are 2
solutions of (2.20) for some initial data $f_{o}$, (4.6) and
property (3.5) of $Q$ give:
\begin{equation*}
\parallel\left(f - g\right)(t)\parallel \, \leq \,2C\parallel f_{o}\parallel\int_{0}^{t}
\parallel\left(f - g\right)(s)\parallel\,ds, \quad \forall t \geq
0.
\end{equation*}
which gives, by Gronwall Lemma, $f = g$, and the uniqueness
follows. \\Now as we indicated, solving the initial value problem
for the relativistic Boltzmann equation is equivalent to solving
the integral equation (2.20). This ends the proof of theorem
4.1$\Box$.\\
\textbf{Remark 4.1:}In a future paper, we study the Boltzmann
equation coupled to the Einstein equations in the case of several
species
of particles with different rest masses.\\
\textbf{Acknowledgements}: The authors thank A. D. Rendall for
helpful comments and suggestions. This work was supported by the
VolkswagenStiftung, Federal Republic of Germany.

\end{document}